\documentclass[11pt,a4paper,oneside]{article}
\usepackage[T1]{fontenc}
\usepackage[ansinew]{inputenc}
\usepackage[english]{babel}
\usepackage{amsfonts}
\usepackage{amsmath}
\usepackage{bm}
\usepackage{array}
\usepackage{amsthm}
\usepackage{amssymb}
\usepackage{graphicx}
\usepackage{braket}
\usepackage{verbatim}
\usepackage[table]{xcolor}
\usepackage{caption}
\usepackage{cite}
\usepackage{textcomp}
\usepackage{url}
\usepackage{hyperref}
\usepackage{mathtools}

\usepackage{subcaption}
\hypersetup{
    colorlinks=true,
    linkcolor=black,
    citecolor=black
    }
\usepackage{multicol}
\newcommand{\be}{\begin{eqnarray}}
\newcommand{\ee}{\end{eqnarray}}
\newcommand{\dd}{{\rm d}}
\newcommand{\ber}[2]{\text{ber}_{#1}\left(#2\right)}
\newcommand{\bei}[2]{\text{bei}_{#1}\left(#2\right)}

\definecolor{pinegreen}{rgb}{0.0, 0.47, 0.44}

\theoremstyle{definition}

\theoremstyle{remark}

\title{\bf On linear waves with memory in a Bessel-like medium}
\author{
A.~Giusti$^{ab}$\thanks{Email: andrea.giusti9@unibo.it},
$\ $
I.~Colombaro$^{c}$\thanks{Email: Ivano.Colombaro@unibz.it},
$\ $
A.~Mentrelli$^{bd}$\thanks{E-mail: andrea.mentrelli@unibo.it}
\\
\\
$^a${\em DIFA \& ${\cal AM}^2$, University of Bologna, 40126 Bologna, Italy}
\\
\\
$^b${\em I.N.F.N., Sezione di Bologna, I.S.~FLAG, 40127 Bologna, Italy}
\\
\\
$^c${\em Faculty of Engineering, Free University of Bozen-Bolzano, 39100 Bolzano, Italy}
\\
\\
$^d${\em Department of Mathematics \& ${\cal AM}^2$, University of Bologna, 40123 Bologna, Italy}
}
\begin{document}
\maketitle
\begin{abstract}
\noindent We discuss the propagation of harmonic and transient waves for systems governed by a wave equation with memory whose integral kernel involves ratios of modified Bessel functions of the first kind in the Laplace domain. In particular, the investigation of transient waves is carried out by means of a fully numerical approach based on the Talbot method for the numerical inversion of Laplace transforms.
\end{abstract}

\newpage

\section{Introduction}
\label{sec:intro}

Ratios of Bessel functions of contiguous order $\nu>-1$ have shown up in several different settings, from hemodynamics \cite{MainardiICMMB82,MainardiICMMB83,GiustiMainardi2016} to viscoelasticity \cite{CGM2017} and electrical systems \cite{GM-infinite-2016}.
Within these physical scenarios one can typically derive wave equations of the form
\be 
\label{eq:WaveEq-general}
\frac{\partial^2 Y}{\partial {t}^2} 
- c^2 \, \big[ 1 - \Phi_{\nu} (t) \ast \big]\,\frac{\partial^2 Y}{\partial {x}^2}
= 0 \, ,
\ee
with $Y = Y(t,x)$, $c$ the wave-front velocity, $\Phi _{\nu}(t)$ being the memory kernel taking the general form
$$
\widetilde{\Phi}_{\nu} (s) = \frac{2 (\nu + 1)}{\sqrt{s \tau}}
\frac{I_{\nu+1} \left(\sqrt{s \tau} \right)}{I_{\nu} \left(\sqrt{s \tau} \right)}
$$
in the Laplace domain (with $\tau$ the relaxation time of the system), $I_\nu (z)$ being the modified Bessel function of the first kind, and where $\ast$ denotes the Laplace convolution, i.e. 
$$
(f \ast g) (t) := \int _0 ^t \, f(t-\xi) \, g(\xi) \, \dd \xi \, .
$$

The problem of the propagation of transient waves for the wave equation \eqref{eq:WaveEq-general} has been preliminary investigated in \cite{CGM-ZAMP} by taking advantage of a semi-analytical method known as the Buchen-Mainardi algorithm (see \cite{CGM-WaveMotion2017, Mainardi2010book} and references therein), which is however reliable only close to the wave-front.

In this work we expand on the investigation of the propagation of waves for systems described by wave equations of the form \eqref{eq:WaveEq-general}, taking as reference the case with $\nu = 0$ (i.e. the model originally proposed in \cite{MainardiICMMB82} and later revisited in \cite{GiustiMainardi2016}). In other words, we shall investigate the features of the following wave equation with memory
\be 
\label{eq:WaveEq}
\frac{\partial^2 Y}{\partial {t}^2} 
- c^2 \, \big[ 1 - \Phi (t) \ast \big]\,\frac{\partial^2 Y}{\partial {x}^2}
= 0 \quad \mbox{with} \quad 
\widetilde{\Phi} (s) = \frac{2}{\sqrt{s \tau}}
\frac{I_{1} \left(\sqrt{s \tau} \right)}{I_{0} \left(\sqrt{s \tau} \right)} \, . 
\ee
The scope of this work is twofold. First, we discuss the effect that the memory term in \eqref{eq:WaveEq} has on the dispersion and attenuation of harmonic waves {\em via} a detailed study of the spatially-attenuated and temporally-periodic branch of the associated dispersion relation. Second, we build upon the analysis in \cite{CGM-ZAMP} by numerically computing the step-response of a system described by Eq.~\eqref{eq:WaveEq}, extending the result past the wave-front approximation. This will be achieved by employing the Talbot method for numerically for numerically inverting Laplace transforms (see \cite{Talbot}, and also \cite{GGM2021}).

\section{Dispersion and Attenuation}
\label{sec:disp}
Let us observe that Eq.~\eqref{eq:WaveEq} can be understood as a reformulation of
\be 
\label{eq:WaveEq-1}
\frac{\partial^2 Y}{\partial {t}^2} 
- c^2 \, \big[ 1 - \Phi (t) \star \big]\,\frac{\partial^2 Y}{\partial {x}^2}
= 0 \quad \mbox{with} \quad 
\widehat{\Phi} (\omega) = \frac{2}{\sqrt{\imath \omega \tau}}
\frac{I_{1} \left(\sqrt{\imath \omega \tau} \right)}{I_{0} \left(\sqrt{\imath \omega \tau} \right)} \, ,
\ee
for functions $Y(t,x)$ that are {\em causal}, where $\star$ denotes the Fourier convolution
$$
(f \star g) (t) := \int _{-\infty} ^{+\infty} \, f(t-\xi) \, g(\xi) \, \dd \xi \, ,
$$
and $\widehat{\Phi} (\omega)$ represents the Fourier transform of $\Phi (t)$. Note that $\Phi (t)$, obtained as the inverse Fourier transform of $\widehat{\Phi} (\omega)$, is a causal function since $\widehat{\Phi} (\omega)$ only has simple poles of the form $\omega = \imath y$ with $y > 0$ (see e.g. \cite{GiustiMainardi2016} and \cite{AGThesis}).

We can now investigate the dispersion and attenuation properties for the wave equation~\eqref{eq:WaveEq-1} using standard dispersion analysis techniques (see e.g. \cite{Mainardi2010book}). We begin by computing the dispersion relation for \eqref{eq:WaveEq-1}. Consider the standard {\em ansatz} for harmonic wave solutions, i.e.
\begin{equation}
\label{eq:ansatz}
{Y} (t, x) = {A} \, {\rm e}^{\imath \left(\omega t - k x \right)} \, ,
\end{equation}
for Eq.~\eqref{eq:WaveEq-1}, with ${A} \in \mathbb{R}^+$ being a constant, which yields the {\em dispersion relation}
\begin{equation}
\label{eq:DR}
(\imath \omega)^2 
+ k^2 c^2 \, \big[ 1 - \widehat{\Phi} (\omega) \big]
= 0 \, .
\end{equation}
In general, the dispersion relation \eqref{eq:DR} is a constraint for $\omega , k \in \mathbb{C}$. However, different choices of boundary conditions lead to specific restrictions on $k$ and $\omega$, see e.g. \cite{ZhangOstoja}. 

Restricting this discussion to harmonic waves that are attenuated in space and periodic in time\footnote{Following the terminology in \cite{ZhangOstoja}, such solutions are known as spatially-attenuated and temporally-periodic (SATP) harmonic waves.}, i.e. to the branch of solutions to \eqref{eq:DR} such that $\omega \in \mathbb{R}$ and $k \in \mathbb{C}$. Observe, in fact, that if 
\begin{equation}
\label{eq:branch}
\omega \in \mathbb{R} \quad \mbox{and} \quad k = \kappa - \imath \, \delta_{\rm att} \, ,
\end{equation}
with $\kappa,\delta_{\rm att} \in \mathbb{R}$, then
$$
{Y} (t, x) = {A} \, {\rm e}^{\imath \left(\omega t - k x \right)} = 
{A} \, {\rm e}^{-\delta_{\rm att} \, x} \, {\rm e}^{\imath \left(\omega t - \kappa x \right)} \, ,
$$
thus explicitly displaying the periodicity in time and attenuation in space (if $\delta_{\rm att} > 0$). For this reason we have that, for this branch of solutions, the quantity
$$ 
\delta_{\rm att} := - {\rm Im} (k)  $$ is referred to as the {\em attenuation factor}. Furthermore, for SATP harmonic waves we can introduce the notion of {\em phase velocity} as 
\be
\label{eq:def-phasev}
v_{\rm p} := \frac{\omega}{\kappa} \equiv \frac{\omega}{{\rm Re} (k)} \, .
\ee
Now, since we are interested in SATP harmonic waves, we should first solve Eq.~\eqref{eq:DR} for $k^2 = [k (\omega)]^2$; this yields
\begin{equation}
\label{eq:dis-inter-1}
    [k (\omega)]^2 = \frac{1}{c^2}\frac{\omega^2}{1 - \widehat{\Phi} (\omega)} =  \frac{\omega^2}{c^2}\frac{I_0 \left(\sqrt{\imath \omega \tau} \right)}{I_2 \left(\sqrt{\imath \omega \tau} \right)}  \, ,
\end{equation}
with $\omega \in \mathbb{R}$, where in the second equality we have taken advantage of the 
recurrence relation (see {\em e.g.}, \cite[9.6.26]{Abramowitz})
$$
I_{\beta - 1} (z) - \frac{2 \beta}{z} I_{\beta} (z) = I_{\beta + 1} (z) 
$$
for modified Bessel functions of the first kind.

\begin{figure}
    \centering
    \begin{subfigure}[b]{6cm}
    \centering
    \includegraphics[width=6cm]{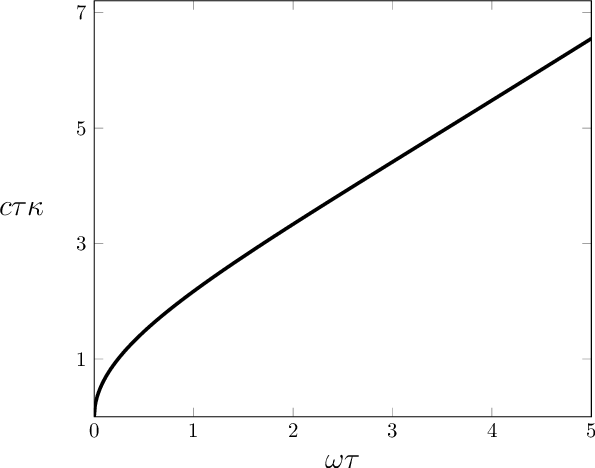}
    \caption{Dispersion law.\label{fig:image1}}
    \end{subfigure}
    \quad
    \begin{subfigure}[b]{6.1cm}
    \centering
    \includegraphics[width=6.1cm]{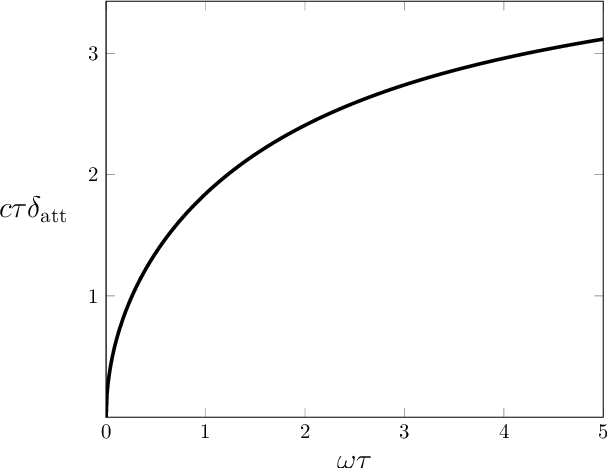}
    \caption{Attenuation factor.\label{fig:image2}}
    \end{subfigure}
    \caption{Non-dimensional dispersion law and attenuation factor as a function of $\omega \tau$. \label{fig:1}}
\end{figure}

Let us observe that 
\be 
{\rm Re} \left[ k^2 (\omega) \right] &\!\!=\!\!& ({\rm Re} \left[ k (\omega) \right])^2 - ({\rm Im} \left[ k (\omega) \right])^2 = \kappa^2 (\omega) -  \delta_{\rm att}^2 (\omega) \, , \label{eq:kd-1} \\
{\rm Im} \left[ k^2 (\omega) \right] &\!\!=\!\!& 2 ({\rm Re} \left[ k (\omega) \right])({\rm Im} \left[ k (\omega) \right]) = - 2 \, \kappa (\omega) \, \delta_{\rm att} (\omega) \, . \label{eq:kd-2}
\ee
Then, recalling the series representations of the {\em Kelvin functions} $\ber{\alpha}{z}$ and $\bei{\alpha}{z}$ \cite[(10.65.1)]{DLMF}, i.e.
\begin{equation}
\label{eq:ber}
\ber{\alpha}{z} := \left(\frac{z}{2}\right)^\alpha  \sum_{k=0}^\infty
\frac{\cos\left[\left( \frac{3\alpha}{4}+\frac{k}{2}\right)\pi\right]}{k!\Gamma(k+\alpha+1)} \left(\frac{z^2}{4}\right)^k \, ,
\end{equation}
\begin{equation}
\label{eq:bei}
\bei{\alpha}{z} := \left(\frac{z}{2}\right)^\alpha  \sum_{m=0}^\infty
\frac{\sin\left[\left( \frac{3\alpha}{4}+\frac{m}{2}\right)\pi\right]}{m!\Gamma(m+\alpha+1)} \left(\frac{z^2}{4}\right)^m \, ,
\end{equation}
and taking advantage of the analysis presented in \cite{CGM2023}, one finds (following extensive yet simple computations, see \cite{AGThesis}) that
\begin{equation}
\label{eq:rek2}
{\rm Re} \left[ k^2 (\omega) \right] =
-\frac{\omega^2}{c^2} \frac{\ber{0}{\sqrt{\omega \tau}}\ber{2}{\sqrt{\omega \tau}} 
+ \bei{0}{\sqrt{\omega \tau}}\bei{2}{\sqrt{\omega \tau}}}{(\ber{2}{\sqrt{\omega \tau}})^2 + (\bei{2}{\sqrt{\omega \tau}})^2} =: A(\omega) \, ,
\end{equation}
\begin{equation}
\label{eq:imk2}
{\rm Im} \left[ k^2 (\omega) \right] =
-\frac{\omega^2}{c^2} 
\frac{\bei{0}{\sqrt{\omega \tau}} \ber{2}{\sqrt{\omega \tau}} - 
\ber{0}{\sqrt{\omega \tau}}\bei{2}{\sqrt{\omega \tau}}}{(\ber{2}{\sqrt{\omega \tau}})^2 + (\bei{2}{\sqrt{\omega \tau}})^2} =: B(\omega) \, .
\end{equation}
Hence, we now have to look for {\em real solutions} of this nonlinear system of equations in $(\kappa, \delta_{\rm att})$ given by the pair of Eqs.~\eqref{eq:kd-1} and \eqref{eq:kd-2} with the prescriptions in Eqs.~\eqref{eq:rek2} and \eqref{eq:imk2}. In other words, we have to find real solutions of the nonlinear system
\be
\left\{
\begin{aligned}
& \kappa^2 - \delta_{\rm att}^2 = A \, , \\
& - 2 \, \kappa \, \delta_{\rm att} = B \, ,
\end{aligned}
\right.
\ee
which admits solutions given by
\be 
\label{eq:syssol}
\left\{
\begin{aligned}
& \kappa = \pm \frac{\sqrt{A-\sqrt{A^2+B^2}}}{\sqrt{2}} \, ,\\
& \delta_{\rm att} = \mp \frac{B}{\sqrt{2} \sqrt{A-\sqrt{A^2+B^2}}} \, ,
\end{aligned}
\right.
\quad \mbox{and} \quad
\left\{
\begin{aligned}
& \kappa = \pm \frac{\sqrt{A+\sqrt{A^2+B^2}}}{\sqrt{2}} \, ,\\
& \delta_{\rm att} = \mp \frac{B}{\sqrt{2} \sqrt{A+\sqrt{A^2+B^2}}} \, . 
\end{aligned}
\right.
\ee
It is now trivial to observe that
$$
A(\omega)-\sqrt{A(\omega)^2+B(\omega) ^2} \leq 0 \quad \mbox{and}\quad A(\omega)+\sqrt{A(\omega)^2+B(\omega) ^2} \geq 0 \, ,
$$
for $\omega\geq 0$. Hence we can only consider the second system in \eqref{eq:syssol} since the first would otherwise lead to complex solutions. Choosing the {\em positive branch} of the solution in the second system in \eqref{eq:syssol}, i.e.
\begin{equation}
\label{eq:dispersion-law}
\kappa (\omega) \equiv {\rm Re} \left[ k (\omega) \right] = \sqrt{\frac{A(\omega)+\sqrt{A(\omega)^2+B(\omega)^2}}{2}} \, ,    
\end{equation}
\be 
\label{eq:attenuation}
\delta_{\rm att} (\omega) \equiv - {\rm Im} \left[ k (\omega) \right] = -
\frac{B(\omega)}{\sqrt{2} \sqrt{A(\omega)+\sqrt{A(\omega)^2+B(\omega)^2}}} \, .
\ee
The behavior of both these quantities as a function of the frequency is depicted, in non-dimensional units, in {\bf Fig.~\ref{fig:1}}.

\section{Phase and Group velocities}
Two quantity of interest that play an important role in determining the dispersion properties of a medium are the so-called {\em phase} and {\em group velocities}. 

The {\em phase velocity} was already defined in Eq.~\eqref{eq:def-phasev} for SATP harmonic waves. Hence, taking advantage of Eq.~\eqref{eq:dispersion-law} one finds that this velocity for the problem at hand reads
\be 
v_{\rm p} (\omega) = \frac{\omega}{\kappa (\omega)} =\sqrt{\frac{2\omega^2}{A(\omega)+\sqrt{A(\omega)^2+B(\omega)^2}}} \, .
\ee
Notably, in analogy with the discussion in \cite{CGM2017}, taking advantage of the asymptotic expansion for modified Bessel functions \cite[(9.7.1)]{Abramowitz} one can easily show that $v_{\rm p} (\omega) \to c$ as $\omega \tau \to \infty$.

The {\em group velocity}, instead, is defined for SATP harmonic waves as
\begin{equation}
v_{\rm g} (\omega) := 
\left[ \frac{{\rm d} \kappa (\omega)}{{\rm d} \omega} \right]^{-1} \, .
\end{equation}
Considering the dispersion law \eqref{eq:dispersion-law} we find that
\be
\frac{{\rm d} \kappa(\omega)}{{\rm d} \omega} =\frac{1}{4\kappa(\omega)}\left[A'(\omega)+\frac{A(\omega)A'(\omega)+B(\omega)B'(\omega)}{\sqrt{A(\omega)^{2}+B(\omega)^{2}}}\right] \, ,
\ee
where the prime compactly denotes the derivative with respect to $\omega$, while $A(\omega)$ and $B(\omega)$ read as in Eqs.~\eqref{eq:rek2} and~\eqref{eq:imk2}. An explicit expression for $v_{\rm g} (\omega)$ can then be computed analytically taking the derivatives of $A(\omega)$ and $B(\omega)$ and exploiting the properties of the Kelvin functions. However, the final result is an extremely long and convoluted expression involving Kelvin functions of arguably limited general interest. Such an expression can be easily obtained explicitly with \textsc{Wolfram Mathematica}\textsuperscript{\textregistered}, but it is arguably of limited use. On the other hand, since we are mostly interested in the plot of the group velocity as a function of $\omega \tau$ we can take advantage of the finite-difference method to numerically approximate the derivative of $\kappa$ with respect to $\omega$ (see e.g. \cite[\textsection 9.10.1]{Quarteroni}). The plots obtained from either of the two procedures are nearly indistinguishable, as one would naturally expect.

\begin{figure}
    \centering
    \includegraphics[width=8cm]{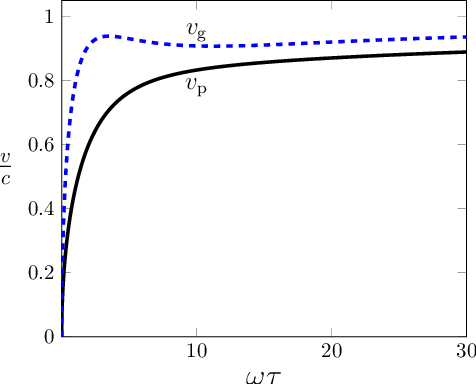}
    \caption{Normalized comparison of the phase (continuous line) and group (dashed line) velocities as a function of $\omega \tau$. \label{fig:2}}
\end{figure}

A comparison of the phase and group velocities is provided in {\bf Fig.~\ref{fig:2}}. 
The group velocity represents the speed at which the modulation of a carrier wave propagates. This suggests that the group velocity represents the speed at which the ``information'' carried by a wave-packet is transmitted, at least in scenarios where this velocity has a well-defined physical meaning \cite{Biot,Brillouin1960,MainardiWM1987,Whitham} (although things might not be as transparent when dealing with dissipative systems \cite{MainardiWM1987,VanGroesenMainardi}).

Numerically it is possible to observe that for our dispersion law \eqref{eq:dispersion-law} one has that $v_{\rm p} (\omega) \lesssim  v_{\rm g} (\omega) \lesssim c$ for $\omega \geq 0$ and that both velocities approach the wave-front velocity $c$ as $\omega \to \infty$. The situation in which both group and phase velocities make physical sense (i.e. they do not exceed $c$) and $v_{\rm p} <  v_{\rm g}$ constitutes a case of {\em anomalous dispersion} \cite{Mainardi2010book}. Unfortunately, however, these result can hardly be proven  rigorously given the complexity of both the expression for the group velocity and the asymptotic properties of Kelvin functions.

\section{Propagation of transient waves}
\label{sec:TW}
In order to investigate the propagation of transient waves for a wave equation of the form \eqref{eq:WaveEq} we examine a problem setup similar to the one described in \cite{CGM-WaveMotion2017}. However, differently from \cite{CGM-WaveMotion2017}, we shall carry out a full numerical treatment of the problem.

Let us assume that our system, whose evolution is governed by Eq.~\eqref{eq:WaveEq}, is quiescent for $t<0$, semi-infinite (i.e. $x\geq0$), and subject to an initial sudden disturbance $Y(t,0) = H(t)$ at $x=0$, with $H(t)$ denoting the Heaviside step function. In other words, we are interested in studying the {\em step response} of the system that can be summarized in the following initial-boundary value problem 
\be
\label{eq:ibvp}
\left\{ 
\begin{aligned}
& Y_{tt} - c^2 \, \big[ 1 - \Phi (t) \ast \big]\, Y_{xx} = 0 \, , &x>0 \, , \,\, t>0 \, ,  \\
& Y (t,0) = H(t) \, ,  &t>0 \, , \qquad\,\quad\\
& Y (0,x) = 0 \, , \,\, Y_t (0,x) = 0 \, , &x > 0 \, .\qquad\,\quad
\end{aligned}
\right.
\ee
where, for the sake of convenience, we adopted the convention that subscripts indicate partial derivatives with respect to the corresponding variables.

Taking the Laplace transform of Eq.~\eqref{eq:ibvp} we have that
\be
\label{eq:ibvp-1}
\left\{ 
\begin{aligned}
& s^2 \, \widetilde{Y} - c^2 \, \big[ 1 - \widetilde{\Phi} (s) \big] \widetilde{Y}_{xx} = 0 \, , &x>0 \, ,   \\
& \widetilde{Y} (s,0) = 1/s  \, ,
\end{aligned}
\right.
\ee
where $\widetilde{Y} \equiv \widetilde{Y} (s,x)$ denotes the Laplace transform of ${Y} (t,x)$ with respect to time. This system can be further rewritten as
\be
\label{eq:ibvp-2}
\left\{ 
\begin{aligned}
& \widetilde{Y}_{xx} - \mu ^2 (s) \, \widetilde{Y} = 0 \, , &x>0 \, ,   \\
& \widetilde{Y} (s,0) = 1/s  \, ,
\end{aligned}
\right.
\ee
with 
\be
\label{eq:mu}
\mu ^2 (s) := \frac{s^{2}}{c^{2}\,\big[1-\widetilde{\Phi}(s)\big]} = \frac{s^2}{c^2} \left[\frac{I_0 \left(\sqrt{s \tau} \right)}{I_2 \left(\sqrt{s \tau} \right)} \right] \, .
\ee
The general solution of \eqref{eq:ibvp-2}$_{1}$ is given by
\be 
\widetilde{Y} (s,x) = C_1 \, {\rm e}^{\mu (s) \, x} + C_2 \, {\rm e}^{-\mu (s) \, x} \, .
\ee
Requiring a bounded solution as $z \to +\infty$ and imposing the boundary condition $\widetilde{Y} (s,0) = 1/s$ yields
\be
\label{eq:LTtoinvert}
\widetilde{Y}(s,x) = \frac{1}{s} \, {\rm e}^{-\mu (s) \, x} \, .
\ee
We shall now compute the inverse Laplace transform of $\widetilde{Y} (s,x)$ in order to obtain the space-time profile of the transient wave solution, namely
\be 
\label{eq:sol-InvLT}
Y(t,x) = 
\frac{1}{2\pi \imath} \int_{{\cal B}{\rm r}} {\rm e}^{st} \, \widetilde{Y}(s,x) \, \dd s
=\frac{1}{2\pi \imath} \int_{{\cal B}{\rm r}} \frac{1}{s} \, {\rm e}^{st -\mu (s) x} \, \dd s \, ,
\ee
where ${\cal B}{\rm r} := \{ s \in \mathbb{C} \, | \, {\rm Re} (s) = \gamma \}$ denotes the Bromwich path (see e.g. \cite{Davies}), where $\gamma \geq 0$ is {\em greater} than the real part of all singularities of $\widetilde{Y}(s,x)$.

In this work we approach the computation of \eqref{eq:sol-InvLT} numerically. To this end we employ a version of the {\em Talbot method} \cite{Talbot} for numerically approximating (under suitable conditions, see \cite{Davies}) the result of the Bromwich integral. The underlying idea of this approach is to replace (by taking advantage of the Cauchy theorem) the integral over Bromwich line with one equivalent and deformed contour ${\cal C}:=\{ s(u) \, | \, u \in \mathbb{R}\}$ suitably chosen to encompass all singularities of $\widetilde{Y}(s,x)$ and placing the limits in the left half-plane, where ${\rm e}^{st}$ is exponentially damped for $t > 0$ (an example is given by the parabolic contour of the modified Talbot method, see e.g. \cite{GGM2021}). More concretely, this implies
\begin{equation}
\begin{split}
Y(t,x) &= 
\lim_{R \to \infty} \frac{1}{2\pi \imath} \int_{\gamma - \imath R} ^{\gamma + \imath R} {\rm e}^{st} \, \widetilde{Y}(s,x) \, d s = 
\frac{1}{2\pi \imath} \int_{{\cal C}} {\rm e}^{st} \widetilde{Y}(s,x) \, d s\\ 
&=
\frac{1}{2\pi \imath} \int_{-\infty} ^{+\infty} {\rm e}^{s (u) \, t} \, 
\widetilde{Y}(s (u),x) \, s'(u) \, d u \, , 
\end{split}
\end{equation}
for $t>0$, then one proceeds with the discretization of the latter integral over the real line by taking advantage of standard quadrature rules. Deforming the Bromwich contour as described in the Talbot method, when possible, alleviates issues related to round-off errors due to the exponential suppression of the integrand in the half-plane $\{ s \in \mathbb{C} \, | \, {\rm Re} (s) \leq \gamma \}$ for $t>0$.

Following a discussion analogous to that of \cite{CGM2017}, one can infer that the singularities of \eqref{eq:LTtoinvert} are located on the negative real axis (including $s=0$), hence the application of the Talbot method can be carried out as described in \cite{GGM2021} or \cite{Abate2004}.

\begin{figure}
    \centering
    \includegraphics[width=8cm]{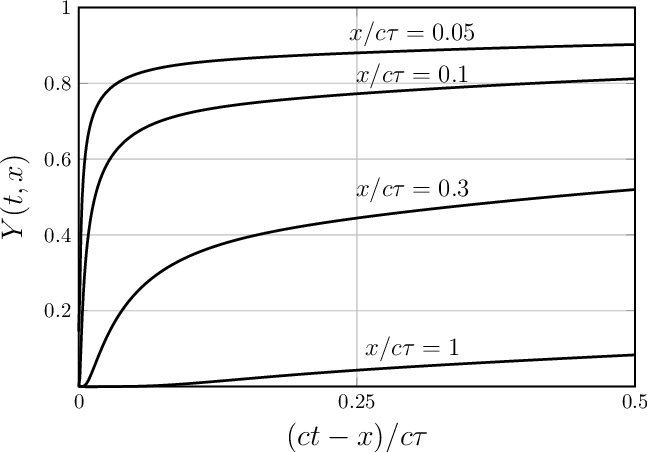}
    \caption{Solution of the system \eqref{eq:ibvp} at different locations $x>0$ (normalized with respect to the characteristic length $c \tau$) as a function of $(ct - x)/c \tau$. \label{fig:3}}
\end{figure}

The numerical results for the step response for a system described by the wave equation with memory \eqref{eq:WaveEq-1}, obtained by means of the numerical approximation of the inverse Laplace transform in \eqref{eq:sol-InvLT} via the method discussed above, are presented in {\bf Fig.~\ref{fig:3}}.

\newpage

\section{Conclusions}
\label{sec:conc}
In this work we have investigated dispersive properties and the propagation of transient waves for Bessel-like systems governed by a wave equation with memory of the form \eqref{eq:WaveEq}.

First, we computed the dispersion relation associated with our wave equation. Then, considering the case 
of spatially-attenuated and temporally-periodic SATP harmonic waves we computed the associated dispersion law \eqref{eq:dispersion-law} and attenuation factor \eqref{eq:attenuation} as functions of the frequency. Using these results we computed the phase and group velocities for SATP harmonics and it was shown numerically that $v_{\rm p} (\omega) <  v_{\rm g} (\omega) \lesssim c$ for $\omega \geq 0$. The fact that both velocities do not exceed the wave-front velocity and that $v_{\rm p} <  v_{\rm g}$ allows us to conclude that the system governed by the wave equation \eqref{eq:WaveEq} exhibits anomalous dispersion.

Second, we investigated the propagation of transient waves for the wave equation \eqref{eq:WaveEq} by computing its associated step response. As opposed to \cite{CGM2017}, we carried out this analysis numerically by taking advantage of the Laplace transform method and the Talbot method for the numerical inversion of Laplace transforms. Our result improves on that of \cite{CGM2017}, which is otherwise reliable only close to the wave-front (i.e. in the limit $(ct-x) \to 0^+$). As one could expect, to obtain reliable results in the transient phase of the plots in {\bf Fig.~\ref{fig:3}} one is forced to consider a large number of nodes in $t$ in the implementation of the Talbot method, this results in a high degree of complexity of the procedure in the transient phase. A way to, perhaps, reduce complexity while maintaining the accuracy of the result would consist in employing some reliable semi-analytical methods to approximate the solution close to the wave-front and then take advantage of the Talbot method form the end of the transient phase onward. This is beyond the scope of this work and is left for future investigation.

\section*{Acknowledgments}
The authors are grateful to G. Dimarco for stimulating discussions. A.G. is supported by the Italian Ministry of Universities and Research (MUR) through the grant ``BACHQ'' (grant no. J33C24003220006) and by the INFN grant FLAG. A.M. is partially supported by MUR under the PRIN2022 PNRR project n. P2022P5R22A. This work has been carried out in the framework of activities of the National Group of Mathematical Physics (GNFM, INdAM).

\section*{Data Availability Statement}
This manuscript has no associated data.
\section*{Code Availability Statement}
This work takes advantage of codes that are already publicly available.

\newpage


\end{document}